# Voltage-Induced Inertial Domain Wall Motion in an Antiferromagnetic Nanowire

Fa Chen[1], Zhendong Zhang[1,#], Wei Luo[1], Xiaofei Yang[1], Long You[1], Yue Zhang[1,*]

1. School of Optical and Electronic Information, Huazhong University of Science and Technology, Wuhan, 430074, PR China.

*E-mail:* yue-zhang@hust.edu.cn (Yue Zhang)

#: The author who has the same contribution with Fa Chen

**Abstract**

Racetrack memory based on magnetic domain walls (DWs) motion exhibits advantages of small volume and high reading speed. When compared to current-induced DW motion, voltage-induced DW motion exhibits lower dissipation. On the other hand, the DW in an antiferromagnet (AFM) moves at a high velocity with weak stray field. In this work, the AFM DW motion induced by a gradient of magnetic anisotropy energy under a voltage pulse has been investigated in theory. The dynamics equation for the DW motion was derived. The solution indicates that the DW velocity is higher than 100 m/s, and because of inertia, the DW is able to keep moving at a speed of around 100 m/s for several nano seconds after turning off the voltage in a period of pulse. The mechanism for this DW inertia is explained based on the Lagrangian route. On the other hand, a spin wave is emitted while the DW is moving, yet the DW is still able to move at an ever increasing velocity with enlarging DW width. This indicates energy loss from emission of spin wave is less than the energy gain from the effective field of the gradient of anisotropy energy.

Motion of magnetic domain walls (DWs) in nanowires is vital to development of novel magnetic memory devices with small volume and high reading speed, such as racetrack memory in which digital information is stored in magnetic domains and read based on DW motion [1]. When compared to DW is induced to move by an electrical current due to spin-transfer-torque (STT) [1-5] or spin-orbit-torque effect [6-9], DW motion triggered by other forces, such as voltage/electrical field [10-12], spin wave [13, 14], and acoustic wave [15, 16], exhibits much lower dissipation. In voltage-induced-DW-motion (VIDWM), a magnetic nanowire is deposited on a wedge-shape insulating medium, and an external voltage generates a gradient of magnetic anisotropy energy that triggers the DW to move [17, 18] (Fig. 1).

In the past, researches were concentrated on DW motion in a ferromagnetic (FM) nanowire. In recent years, more attention is paid on special dynamical behaviors of magnetic textures in other magnetic media, such as antiferromagnet (AFM) [19-23], synthetic antiferromagnet (SAF) [24-26], ferrimagnet [27-30], and magnetic frustrate [31-34]. Typical examples in these cases include ultra-high velocity and depression of Walker breakdown for coupled DWs in an SAF [24, 25], relativistic-like width contraction for an AFM DW, and peculiar AFM DW-magneton interaction [19-21, 23].

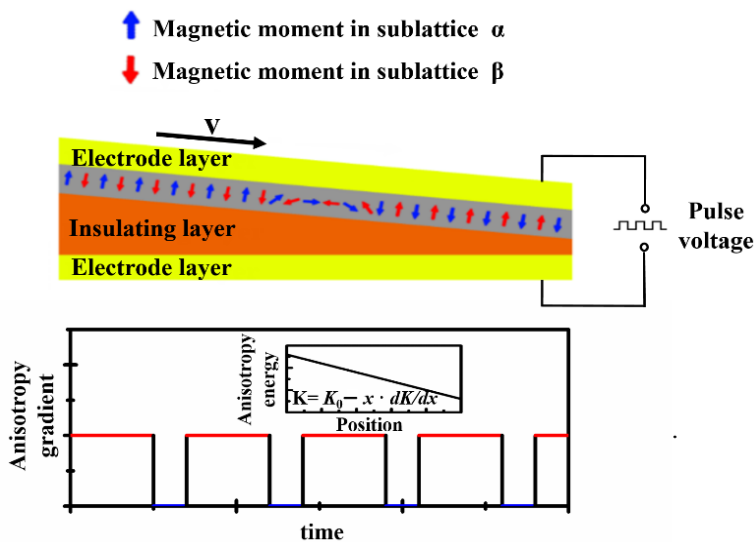

**Figure 1. Schematic of the principle of pulsed voltage induced DW motion in an AFM system: AFM DW is induced to move under a pulsed voltage that generates a coordinate dependent magnetic anisotropy energy with a pulsed gradient of anisotropy constant.**

Very recently, electrical manipulation of AFM properties has attracted attention. For example, Néel vector in an AFM material can be induced to rotate under an electric field [35, 36]. Wen et al also reported their theoretical analysis about VIDWM in an AFM system based on the assumption of DW motion at a *constant* velocity [37]. However, real DW motion is usually more complicated than simple uniform linear motion. In theory, DW motion appears to be analogous to motion of an object with mass, and quantities such as mass, momentum, and inertia, are exploited to analyze DW motion. For example, an effective mass is proposed based on variation of kinetic energy of DW driven under a magnetic field [38]. In STT-induced DW motion, DW inertia due to transfer of momentum from propagating electrons to DW has been predicted theoretically and confirmed experimentally [39-41]. Similarly, AFM DW also exhibits inertia because of the secondary differential dynamics equation that depicts DW motion [19, 20]. Because of inertia, AFM DW motion should be more complicated than uniform linear motion, especially for DW motion induced by a short pulse used commonly in application.

In this letter, we report our theoretic calculation about DW inertial motion at a high velocity (around 100 m/s) in an AFM nanowire driven by a gradient of anisotropy energy induced by constant or pulsed voltage (Fig. 1). The dynamics equation depicting the VIDWM is firstly derived from Landau-Liftshiz-Gilbert (LLG) equation based on the model of an AFM nanowire with two FM sublattices ($\alpha$ and $\beta$) (Fig. 1). The AFM magnetic moments are in a 3D medium but we only consider AFM exchange coupling along the length direction of nanowire. The magnetization in the sublattice $\alpha$ ($\beta$) is expressed as $\vec{M}_{\alpha(\beta)} = M_S \vec{m}_{\alpha(\beta)}$, and $M_S$ is the saturation magnetization of each sublattice and $\vec{m}_{\alpha(\beta)}$ is the unit vector for the orientation of magnetization.

The Hamiltonian of this AFM model is [19]:

$$H_{1D} = \frac{L_y L_z}{\Delta_j \Delta_k} [J \sum_{\langle \alpha, \beta \rangle}^{N,N} \vec{m}_\alpha \cdot \vec{m}_\beta - K_z (\sum_\alpha^N m_{\alpha z}^2 + \sum_\beta^N m_{\beta z}^2)] \qquad (1).$$

In Eq. (1), $J$ is the exchange constant for the nearest AFM exchange coupling. $K_z$ is the uniaxial anisotropy constant. $L_y$, $L_z$, $\Delta_j$, and $\Delta_k$ are the width, thickness, and cell size in the width and thickness direction, respectively. The uniform magnetization and staggered order parameters for the $i$-th cell $\vec{m}_i$ and $\vec{n}_i$ are introduced as:

$$\vec{m}_i = (\vec{m}_\alpha)_i + (\vec{m}_\beta)_i \qquad (2);$$

$$\vec{n}_i = [(\vec{M}_\alpha)_i - (\vec{M}_\beta)_i]/l \qquad (3).$$

Here $l$ is the absolute value of $(\vec{M}_\alpha)_i - (\vec{M}_\beta)_i$ and close to $2M_S$ for strong exchange coupling.

Under continuum approximation and neglecting edge magnetization and parity-breaking term [19], Eq. (3) is converted into:

$$H_{1D} \approx \int_v (\frac{a}{2} m^2 + \frac{A}{2} (\frac{d\vec{n}}{dx})^2 - K n_z^2) d\tau \qquad (4).$$

In Eq. 4, $a$ is a homogeneous exchange energy ($a = 5J/V_{cell}$); $A$ is an exchange stiffness constant ($A = J\Delta_i^2/V_{cell}$) and $K = K_z/V_{cell}$. Here $\Delta_i$ and $V_{cell}$ represent cell size in the length direction and the volume of AFM unit cell, respectively.

Based on Eq. (4), the effective magnetic fields for $\vec{m}$ and $\vec{n}$ were derived using the calculus of variations [20]:

$$\vec{H}_m = -\frac{\delta H_{1D}}{\mu_0 \delta \vec{m}} = -\frac{a\vec{m}}{\mu_0} \tag{5},$$

and

$$\vec{H}_n = -\frac{\delta H_{1D}}{\mu_0 \delta \vec{n}} = \frac{1}{\mu_0}(A\frac{d^2\vec{n}}{dx^2} + 2Kn_z\vec{e}_z) \tag{6}.$$

Here, $\mu_0$ is the permittivity of vacuum.

Starting from the LLG equation, we have formulated the dynamics equations for $\vec{m}$ and $\vec{n}$ under strong exchange coupling:

$$\partial_t \vec{m} = -\frac{\gamma}{M_S}\vec{n}\times\vec{H}_n + \frac{G_2}{M_S}\vec{n}\times\partial_t\vec{n} \tag{7};$$

$$\partial_t \vec{n} = -\frac{\gamma}{M_S}\vec{n}\times\vec{H}_m + G_1 M_S \vec{n}\times\partial_t\vec{m} \tag{8}.$$

Here $G_1$ and $G_2$ are the effective damping parameters for $\vec{n}$ and $\vec{m}$ ($G_1 = \alpha/l$, and $G_2 = \alpha l$). Combing Eqs. (7) and (8) under linear approximation and considering $G_1 \ll G_2$, we have derived the dynamics equations for $\vec{n}$:

$$\ddot{\vec{n}} + \frac{G_2}{fM_S}\dot{\vec{n}} = \frac{\gamma}{fM_S}\vec{H}_n \tag{9}.$$

Here $f = \mu_0 M_S(1+G_1 G_2)/a\gamma$, and $\vec{n} = \vec{n}(x,\{q_1(t), q_2(t)\})$ with $q_1$ and $q_2$ representing two collective coordinates of DW. Multiplying $\frac{\partial \vec{n}}{\partial q_1}$ and $\frac{\partial \vec{n}}{\partial q_2}$ at the two sides of Eq. (9) and integrating them over the nanowire, one can get the following equation:

$$\vec{M}\cdot\ddot{\vec{q}} + \frac{G_2}{fM_S}\vec{M}\cdot\dot{\vec{q}} = \vec{F} \tag{10}.$$

In Eq. (10), $\vec{M}$ is the tensor of effective mass with its component $M_{ij} = \int_{-\infty}^{+\infty}\frac{\partial \vec{n}}{\partial q_i}\cdot\frac{\partial \vec{n}}{\partial q_j}dx$, and

$$F_i = \frac{\gamma}{fM_S}\int_{-\infty}^{+\infty}\vec{H}_n\cdot\frac{\partial \vec{n}}{\partial q_i}dx.$$

In a Néel-type DW wall, $q_1 = q$, and $q_2 = \phi$, and $\vec{n} = (\sin\theta\cos\phi, \sin\theta\sin\phi, \cos\theta)$, in which $\theta = 2\arctan\{\exp[(x-q)/\lambda]\}$ with $\lambda = \sqrt{A/K}$ for DW width [19, 20]. Here $K = K_0 - (dK/dx)x$ [22], and we neglect the variation of DW width. Finally, the dynamics equation for $q$ was derived:

$$\ddot{q} + \frac{G_2}{fM_S}\dot{q} = \frac{\gamma\lambda^2}{\mu_0 fM_S}\frac{dK}{dx} \quad (11).$$

Eq. (11) is solved based on the parameters of NiO [20, 42]: $K_0$, $M_S$, $d$, $\gamma$, and $\mu_0$ are $3.32 \times 10^5$ J/m$^3$, $4.25 \times 10^5$ A/m, $4.2 \times 10^{-10}$ m, $2.21 \times 10^5$ m/A·s, and $4\pi \times 10^{-7}$ N/A$^2$, respectively. Here $M_S$ is estimated from the magnetic moment per sublattice ($1.7\mu_B$ with $\mu_B$ representing the Bohr magneton). $A$, $\alpha$, and $dK/dx$ are in the range of $3 \times 10^{-13}$ J/m ~ $9 \times 10^{-13}$ J/m, $1 \times 10^{-4}$ ~ $8 \times 10^{-4}$, and 0 GJ/m$^4$ ~ 300 GJ/m$^4$.

VIDWM has also been investigated numerically using the micromagnetic simulation software Object-Oriented Micromagnetic Framework (OOMMF). The track dimension is 3000 nm (length) × 1 nm (width) × 0.5 nm (thickness), and the cell dimension is 0.5 nm × 0.5 nm × 0.5 nm.

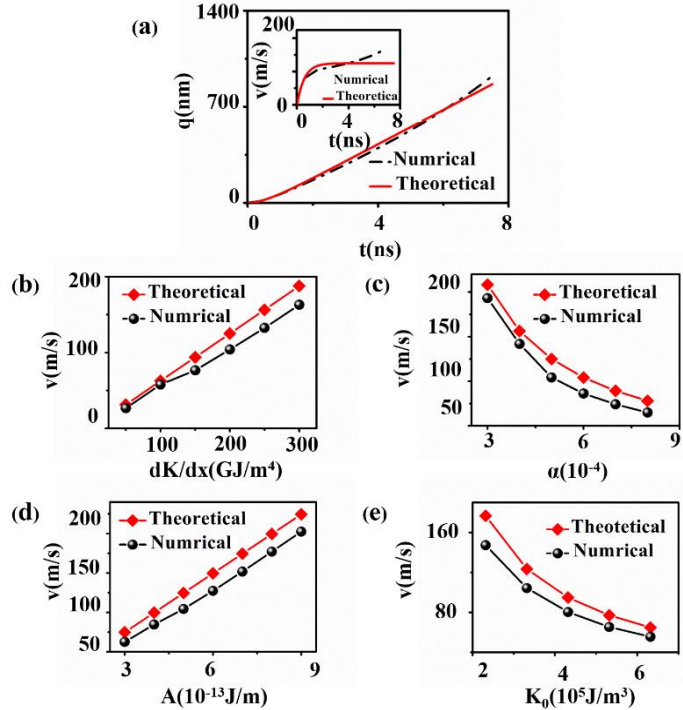

**Figure 2. (a). Temporal displacement of AFM DW under a 200-GJ/m$^4$ d$K$/d$x$ (Inset: Temporal DW velocity); (b) ~ (e). Theoretical stable DW velocity (red) and the DW velocity in the turning point of the simulated velocity/time curves (black) as a function of d$K$/d$x$, damping coefficient**

$\alpha$, exchange stiffness constant $A$, and the intercept $K_0$ in the $x$ dependence of $K$

Initially, Eq. (11) is solved under the parameters d$K$/d$x$ = 200 GJ/m$^4$, $\alpha$ = 5 × 10$^{-4}$, and $A$ = 5 × 10$^{-13}$ J/m. The result (red solid line in Fig. 2(a)) indicates that DW moves at an increasing velocity in the initial 2 ns, and the velocity reaches a stable value (around 120 m/s). The magnitude of DW velocity is the same with that of voltage-induced skyrmion motion [22]. The result of numerical simulation (the black dashed line) is very close to the solution of Eq. (11) in the first 4 ns. Afterwards, the simulated DW motion becomes faster. This acceleration is attributed to the increase of DW width that is neglected in the theoretical calculation.

Based on the theoretical analysis and numerical simulation, the influence of d$K$/d$x$, $\alpha$, and $A$ on DW velocity was studied in detail. When d$K$/d$x$ increases from 0 to 300 GJ/m$^4$, a theoretical stable DW velocity increases monotonously to around 200 m/s ($\alpha$ and $A$ are 5 × 10$^{-4}$ and 5 × 10$^{-13}$ J/m, respectively.). When $\alpha$ increases from 1 × 10$^{-4}$ to 8 × 10$^{-4}$, the DW velocity decreases from around 200 m/s to about 80 m/s (d$K$/d$x$ and $A$ are 200 GJ/m$^4$ and $A$ = 5 × 10$^{-13}$ J/m, respectively.) (Fig. 2(c)). This indicates low damping is favorable for fast DW motion. Such a low $\alpha$ of around 10$^{-4}$ has been observed in NiO experimentally [42]. On the other hand, the increase of $A$ from 5 × 10$^{-13}$ J/m to 9 × 10$^{-13}$ J/m enhances the DW velocity from 75 m/s to about 230 m/s ($\alpha$ and d$K$/d$x$ are fixed as 5 × 10$^{-4}$ and 200 GJ/m$^4$.) (Fig. 2(d)). Additionally, the increase of the intercept $K_0$ in $K(x)$ reduces the DW velocity (Fig. 2(e)). The DW velocity in the turning point of the simulated velocity/time curve was compared with the theoretic stable velocity. The simulation results are close to that in theory, and the difference between simulation and theory is smaller when d$K$/d$x$ is smaller or $K_0$ is higher. In either case, the relative variation of DW width in DW motion is smaller.

Owing to the secondary differential of $q$, AFM DW motion exhibits inertia, i.e., DW experiences an acceleration/deceleration stage after the voltage is turned on/off. To investigate this inertial DW motion, the constant d$K$/d$x$ was replaced by a pulsed d$K$/d$x$. As a representative example for DW motion driven by a single pulse (d$K$/d$x$ = 200 GJ/m$^4$) with a duration of 50 ns, a gradual acceleration and deceleration for around 20 ns appears soon after the pulse for generating this d$K$/d$x$ is turned on and off (Fig. 3(a)). As to SOT-induced motion of an FM DW, secondary differential equation of central position can be also derived by deleting azimuthal angle in the equation group of collective coordinates [43]. However, since the central position is coherent to the azimuthal angle, the effective mass of an FM DW changes with variation of DW structure in the process of DW motion [43]. On the contrary, in VIDWM for AFM, the components of effective mass tensor include $M_{11} = 2/\lambda$ for $q$ and $M_{22} = 2\lambda$ for $\phi$, and both cross terms $M_{12}$ and $M_{21}$ are zero. This means that $q$ is independent to $\phi$, and the effective mass is a constant.

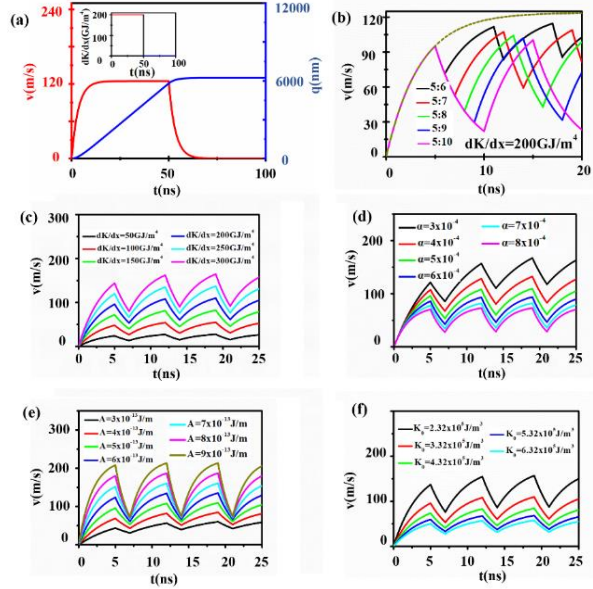

**Figure 3. (a). Temporal displacement and velocity of AFM DW under the gradient of magnetic anisotropy constant d$K$/d$x$ as indicated in the inset, (b). Temporal velocity of AFM DW under a 200-GJ/m$^4$ d$K$/d$x$ and different duty ratios (The dashed line indicates the temporal DW velocity under the constant d$K$/d$x$ of 200 GJ/m$^4$), (c) ~ (f). Temporal velocity of AFM DW driven by gradient of magnetic anisotropy energy as a function of (c) exchange stiffness constant $A$, (d) intercept of the $x$ dependent $K$ ($K_0$), (e) gradient of magnetic anisotropy constant d$K$/d$x$, and (f) damping coefficient $\alpha$.**

The inertia of AFM DW motion is influenced by duty ratio, $A$, $\alpha$, $K_0$, and d$K$/d$x$ (Figs. 3(b) ~ (f)). The temporal velocity of DW motion under a pulsed d$K$/d$x$ with the duty ratio between 5 : 5 (constant d$K$/d$x$ = 200 GJ/m$^4$) and 5 : 10 (5 ns for d$K$/d$x$ = 200 GJ/m$^4$ and 5 ns for d$K$/d$x$ = 0 GJ/m$^4$) shows that the DW keeps moving at an average velocity of around 50 m/s when the zero-d$K$/d$x$ stage lasts as long as 5 ns (Fig. 3(b)). On the other hand, with enhancing $A$, decreasing $K_0$, or increasing d$K$/d$x$, the DW motion under non-zero d$K$/d$x$ and the decaying of DW velocity in the

zero-d$K$/d$x$ stage become faster (Fig. 3(c) ~ (f)). Reducing $\alpha$ enhances the DW velocity under non-zero d$K$/d$x$ and also slows down the decaying of DW velocity under zero d$K$/d$x$. Therefore, low damping is important for the DW motion at a *sustainable* high velocity.

The mechanism of inertia can be understood using the Lagrangian approach. The Lagrangian density $\Omega$ equals $T - V$ with $T$ and $V$ representing kinetic and free energy density, respectively. $V$ is the integrand in Eq. (4), while $T$ is [19]:

$$T = \rho \vec{m} \cdot (\dot{\vec{n}} \times \vec{n}) \tag{12}.$$

Here, $\rho$ is a constant that is proportional to spin angular momentum [19]. For an AFM system with a nonzero damping coefficient, a dissipation density function $R$ is introduced as [44]:

$$R = \tilde{\alpha}(\dot{\vec{n}})^2 \tag{13}.$$

Here $\tilde{\alpha}$ is a phenomenological factor. The Lagrangian ($L$) and Rayleigh dissipation function $F$ were derived via integrating $T$ and $R$ throughout the AFM nanowire, and they satisfy the Lagrange-Rayleigh equation:

$$\frac{d}{dt}(\frac{\partial L}{\partial \dot{q}}) = \frac{\partial L}{\partial q} + \frac{\partial F}{\partial \dot{q}} \tag{14}.$$

Here $\frac{\partial L}{\partial \dot{q}}$, $\frac{\partial L}{\partial q}$, and $\frac{\partial F}{\partial \dot{q}}$ are canonical momentum ($P$), canonical force, and dissipation force, respectively. Combining Eqs. (12) ~ (14), we have derived the dynamics equation of $P$:

$$\frac{dP}{dt} = -\frac{2\tilde{\alpha}}{f(2\rho - af)}P + N\frac{dK}{dx} \tag{15}.$$

In Eq. (15), $P$ and $N$ are:

$$P = \frac{\partial L}{\partial \dot{q}} = 4L_y L_z \frac{f}{\lambda}\left(\rho - \frac{af}{2}\right)\dot{q} + L_y L_z \alpha\lambda\left(f - \frac{\rho}{a}\right)\frac{dK}{dx} \tag{16},$$

$$N = \frac{2L_y L_z \alpha\tilde{\alpha}\lambda}{f(2\rho - af)}\left(f - \frac{\rho}{a}\right) + L_y L_z \frac{2\alpha^2\lambda}{3a}\left(K_0 - \frac{dK}{dx}q\right) - L_y L_z \frac{\alpha^2 A}{3a\lambda} - L_y L_z C_1\lambda \tag{17}.$$

Here, $C_1$ is an integral constant $C_1 = \int_{-\frac{L_x}{2}}^{\frac{L_x}{2}} \tanh^2\varepsilon \, d\varepsilon$.

Eqs. (16) shows that $P$ is composed of two terms: one is similar to mechanical momentum and the other one contributed from gradient of anisotropy energy is analogous to the momentum of magnetic field that induces motion of a charged particle. Under a non-zero d$K$/d$x$, the momentum evolves towards a fixed value gradually based on Eq. (15). While after d$K$/d$x$ is retreated, $P$ decays exponentially. This is consistent with the inertia of DW motion described above.

In addition to inertia, motion of AFM DW also usually experiences Lorentz contraction, i.e., DW width decreases with DW velocity that increases towards a limit value determined by the group velocity of spin wave ($v_g$) [21]. When DW velocity approaches $v_g$, DW energy is released by emitting spin wave, which inhibits further increase in DW velocity. In previous analysis, we did not consider variation of DW width. This is reasonable under a moderate d$K$/d$x$. When the DW is driven under a high d$K$/d$x$, variation of DW width is not negligible. In theory, DW motion with acceleration is expected if variation of DW width is also considered [17].

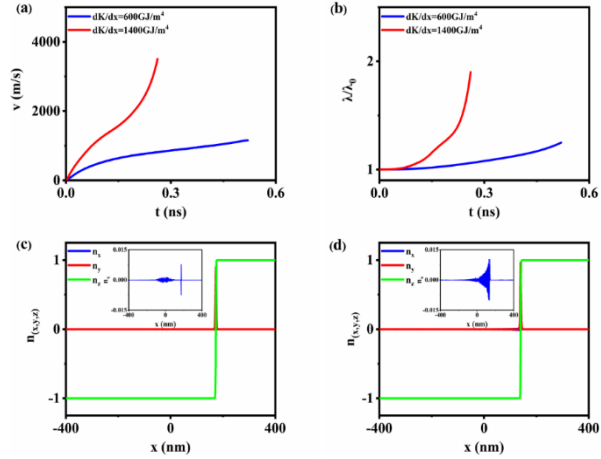

**Figure 4. (a). Temporal DW velocity under two different d$K$/d$x$; (b). evolution of DW width relative to that for a static one under two different d$K$/d$x$; (c). $x$, $y$, $z$ components of Néel vector for a moving DW under 600-GJ/m$^4$ d$K$/d$x$ at a velocity of around 800 m/s; (d). $x$, $y$, $z$ components of Néel vector for a moving DW under 1400-GJ/m$^4$ d$K$/d$x$ at a velocity of around 1500 m/s.**

Using micromagnetic simulation, we have compared the temporal DW velocity and DW width under different d$K$/d$x$ (Fig. 4(a) and (b)). When d$K$/d$x$ = 600 GJ/m$^4$, both DW velocity and DW width increase slowly. While when d$K$/d$x$ = 1400 GJ/m$^4$, DW velocity increases drastically with a turning point appearing in the DW velocity ~ time curve, which is accompanied with obvious widening of DW. On the other hand, it is noticed that a spin wave is also emitted when the DW moves at a high velocity (Fig. 4(c) and (d)). However, this emission of spin wave does not inhibit enhancing of DW velocity. On the contrary, the amplitude of spin wave increases with increasing DW velocity. Since a gradient of anisotropy energy can be converted into an effective magnetic field that becomes stronger with decreasing anisotropy constant [17], sustainable increase in DW velocity indicates the energy gain from this enhancing effective field exceeds the energy loss from the emission of spin wave.

In summary, we have derived the dynamics equation describing DW motion in an AFM nanowire under a moderate gradient of magnetic anisotropy energy. The solution indicates that enhancing the exchange stiffness constant, reducing damping coefficient, or increasing the slope or reducing the intercept of the coordinate dependence of anisotropy energy, increases DW velocity. Because of inertia, the DW is able to keep moving for several nano-seconds after retreating the gradient of magnetic anisotropy energy, and a low damping coefficient is vital to the DW motion at a sustainable high velocity. This DW inertia can be interpreted from the dynamics of canonical momentum that is derived from the Lagrangian approach. On the other hand, even though spin wave is emitted, the DW velocity still keeps increasing under a high gradient of anisotropy energy, showing that the gradient of anisotropy energy offers more energy than the energy loss from emission of spin wave.

**Acknowledgement**

The authors acknowledge financial support from the National Natural Science Foundation of China (No. 51971098) and Science and Technology Department of Hubei Province (No. 2019CFB435).

**References:**

[1]S. S. P. Parkin, M. Hayashi, and L. Thomas, Science **320**, 190 (2008).
[2]M. Hayashi, L. Thomas, R. Moriya, C. Rettner, and S. S. P. Parkin, Science **320**, 209 (2008).
[3]A. Brataas, A. D. Kent, and H. Ohno, Nat. Mater. **11**, 372 (2012).
[4]I. M. Miron, T. Moore, H. Szambolics, L. D. Buda-Prjbeanu, S. Auffret, B. Rodmacq, S. Pizzini, J. Vogel, M. Bonfim, and A. Schuhl, Nat. Mater. **10**, 419 (2011).
[5]A. Thiaville, Y. Nakatani, J. Miltat, and Y. Suzuki, Europhys. Lett. **69**, 990 (2005).
[6]P. P. J. Haazen, E. Murè, J. H. Franken, R. Lavrijsen, H. J. M. Swagten, and B. Koopmans, Nat. Mater. **12**, 299 (2013).
[7]K. S. Ryu, L. Thomas, S. H. Yang, and S. Parkin, Nat. Nanotechnol. **8**, 527 (2013).
[8]S. Emori, U. Bauer, S. M. Ahn, E. Martinez, and G. S. D. Beach, Nat. Mater. **12**, 611 (2013).

[9]S. Emori, E. Martinez, K. J. Lee, H. W. Lee, U. Bauer, S. M. Ahn, P. Agrawal, D. C. Bono, and G. S. D. Beach, Phys. Rev. B **90**, 184427 (2013).

[10]A. J. Schellekens, A. van den Brink, J.H. Franken, H.J.M. Swagten, and B. Koopmans, Nat. Commun. **3**, 847 (2012).

[11]W. Lin, N. Vernier, G. Agnus, K. Garcia, B. Ocker, W. Zhao, E. E. Fullerton, and D. Ravelosona, Nat. Commun. **7**, 13532 (2016).

[12]D. Chiba, M. Kawaguchi, S. Fukami, N. Ishiwata, K. Shimamura, K. Kobayashi, and T. Ono, Nat. Commun. **3**, 888 (2012).

[13]X. Wang, G. Guo, Y. Nie, G. Zhang, and Z. Li, Phys. Rev. B **86**, 054445 (2012).

[14]W. Wang, M. Albert, M. Beg, M. Bisotti, D. Chernyshenko, D. Cortés-Ortuño, I. Hawke, and H. Fangohr, Phys. Rev. Lett. **114**, 087203 (2015).

[15]J. Dean, M. T. Bryan, J. D. Cooper, A. Virbule, J. E. Cunningham, and T. J. Hayward, Appl. Phys. Lett. **107**, 142405 (2015).

[16]W. Edrington, U. Singh, M. A. Dominguez, J. R. Alexander, R. Nepal, and S. Adenwalla, Appl. Phys. Lett. **112**, 052402 (2018).

[17]Y. Zhang, S. Luo, X. Yang, and C. Yang, Sci. Rep. **7**, 2047 (2017).

[18]K. Yamada, S. Murayama, and Y. Nakatani, Appl. Phys. Lett. **108**, 202405 (2016).

[19]E. G. Tveten, T. Müller, J. Linder, and A. Brataas, Phys. Rev. B **93**, 104408 (2019).

[20]E. G. Tveten, A. Qaiumzadeh, and A. Brataas, Phys. Rev. Lett. **112**, 147204 (2014).

[21]T. Shiino, S. H. Oh, P. M. Haney, S. W. Lee, G. Go, B. G. Park, and K. J. Lee, Phys. Rev. Lett. **117**, 087203 (2016).

[22]L. Shen, J. Xia, G. Zhao, X. Zhang, M. Ezawa, O. A. Tretiakov, X. Liu, and Y. Zhou, Phys. Rev. B **98**, 134448 (2018).

[23]E. G. Tveten, A. Qaiumzadeh, O. A. Tretiakov, and A. Brataas, Phys. Rev. Lett. **110**, 127208 (2013).

[24]S. Yang, K. Ryu, and S. Parkin, Nat. Nanotechnol. **10**, 221 (2015).

[25]L. Chen, M. Shen, Y. Peng, X. Liu, W. Luo, X. Yang, L. You, and Y. Zhang, J. Phys. D: Appl. Phys. **52**, 495001 (2019).

[26]C. Ma, X. Zhang, J. Xia, M. Ezawa, W. Jiang, T. Ono, S. N. Piramanayagam, A. Morisako, Y. Zhou, and X. Liu, Nano. Lett. **19**, 353 (2019).

[27]S. Oh, S. K. Kim, J. Xiao, and K. Lee, Phys. Rev. B **100**, 174403 (2019).

[28]E. Haltz, J. Sampaio, R. Weil, Y. Dumont, and A. Mougin, Phys. Rev. B **99**, 104413 (2019).

[29]E. Martínez, V. Raposo, and Ó. Alejos, J. Magn. Magn. Mater. **491**, 165545 (2019).

[30]Y. Hirata, D. H. Kim, S. K. Kim, D. K. Lee, S. H. Oh, D. Y. Kim, T. Nishimura, T. Okuno, Y. Futakawa, H. Yoshikawa, A. Tsukamoto, Y. Tserkovnyak, Y. Shiota, T. Moriyama, S. B. Choe, K. J. Lee, and T. Ono, Nat. Nanotechnol. **14**, 232 (2019).

[31]T. Kurumaji, T. Nakajima, M. Hirschberger, A. Kikkawa, Y. Yamasaki, H. Sagayama, H. Nakao, Y. Taguchi, T. H. Arima, and Y. Tokura, Science **365**, 914 (2019).

[32]K. Karube, J. S. White, D. Morikawa, C. D. Dewhurst, R. Cubitt, A. Kikkawa, X. Yu, Y. Tokunaga, T. Arima, H. M. Rønnow, Y. Tokura, and Y. Taguchi, Sci. Adv. **4**, eaar7043 (2018).

[33]B. Heil, A. Rosch, and J. Masell, Phys. Rev. B **100**, 134424 (2019).

[34]Z. Hou, Q. Zhang, X. Zhang, G. Xu, J. Xia, B. Ding, H. Li, S. Zhang, N. M. Batra, Pmfj Costa, E. Liu, G. Wu, M. Ezawa, X. Liu, Y. Zhou, X. Zhang, and W. Wang, Adv. Mater. **31**, 1904815 (2019).

[35]I. J. Park, T. Lee, P. Das, B. Debnath, G. P. Carman, and R. K. Lake, Appl. Phys. Lett. **114**, 142403 (2019).

[36]Z. Liu, Z. Feng, H. Yan, X. Wang, X. Zhou, P. Qin, H. Guo, R. Yu, and C. Jiang, Adv. Electron. Mater. **5**, 1900176 (2019).

[37]D. L. Wen, Z. Y. Chen, W. H. Li, M. H. Qin, D. Y. Chen, Z. Fan, M. Zeng, X. B. Lu, X. S. Gao, and J. M. Liu, Phys. Rev. Research **2**, 013166 (2020).

[38]G. T. Rado, Phys. Rev. **83**, 821 (1951).

[39]S. E. Barnes and S. Maekawa, Phys. Rev. Lett. **95**, 107204 (2005).

[40]L. Thomas, R. Moriya, C. Rettner, and S. S. P. Parkin, Science **330**, 1810 (2010).

[41]G. Tatara and H. Kohno, Phys. Rev. Lett. **92**, 086601 (2004).

[42]T. Moriyama, K. Hayashi, K. Yamada, M. Shima, Yutak. Ohya, and T. Ono, Phys. Rev. Mater. **3**, 051402 (R) (2019).

[43]J. Torrejon, E. Martinez, and M. Hayashi, Nat. Commun. **7**, 13533 (2016)

[44]R. Cheng and Q. Niu, Phys. Rev. B **88**, 081105 (2014).